%% file: main.tex
\documentclass[acmsmall]{acmart}

\setcopyright{rightsretained}
\acmJournal{TOMM}
\acmYear{2024}
\acmVolume{1}
\acmNumber{1}
\acmArticle{1}
\acmMonth{1}
\acmDOI{10.1145/3678881}

\usepackage{cleveref}
\usepackage{subcaption}
\usepackage{listings}

\usepackage{numprint}

\usepackage[inline]{enumitem}
\usepackage{tikz}
\usetikzlibrary{shapes,shapes.geometric}
\usetikzlibrary{arrows,arrows.meta}
\usetikzlibrary{automata,calc,positioning,matrix,fit}
\usetikzlibrary{decorations.pathmorphing, decorations.pathreplacing, decorations.shapes}
\usepackage{tikzscale}

\begin{document} 

%%
%%
%% Overleaf sharing link: https://www.overleaf.com/1957212576gqxrxwhxjsjs#b8072c
%%
%%

\title{Performance Evaluation in Multimedia Retrieval}

\author{Loris Sauter}
\email{loris.sauter@unibas.ch}
\orcid{0000-0001-8046-0362}
\affiliation{%
  \institution{University of Basel}
  \city{Basel}
  \country{Switzerland}
}

\author{Ralph Gasser}
\email{ralph.gasser@unibas.ch}
\orcid{0000-0002-3016-1396}
\affiliation{%
  \institution{University of Basel}
  \city{Basel}
  \country{Switzerland}
}

\author{Heiko Schuldt}
\email{heiko.schuldt@unibas.ch}
\orcid{0000-0001-9865-6371}
\affiliation{%
  \institution{University of Basel}
  \city{Basel}
  \country{Switzerland}
}

\author{Abraham Bernstein}
\email{bernstein@ifi.uzh.ch}
\orcid{0000-0002-0128-4602}
\affiliation{%
  \institution{University of Zurich}
  \city{Zurich}
  \country{Switzerland}
}

\author{Luca Rossetto}
\email{rossetto@ifi.uzh.ch}
\orcid{0000-0002-5389-9465}
\affiliation{%
  \institution{University of Zurich}
  \city{Zurich}
  \country{Switzerland}
}

\begin{abstract}

Performance evaluation in multimedia retrieval, as in the information retrieval domain at large, relies heavily on retrieval experiments, employing a broad range of techniques and metrics.
These can involve human-in-the-loop and machine-only settings for the retrieval process itself and the subsequent verification of results.
Such experiments can be elaborate and use-case-specific, which can make them difficult to compare or replicate.
In this paper, we present a formal model to express all relevant aspects of such retrieval experiments, as well as a flexible open-source evaluation infrastructure that implements the model.
These contributions intend to make a step towards lowering the hurdles for conducting retrieval experiments and improving their reproducibility.

\end{abstract}



\maketitle

\section{Introduction}

Engaging in experimentation stands as the paramount method for substantiating or refuting hypotheses, thereby pushing the boundaries of human knowledge. Access to precise tools, enabling the formulation, documentation, and reproducibility of experiments, therefore stands as an indispensable element of rigorous scientific inquiry.

While this principle holds true across disciplines, it poses a particularly distinctive challenge in multimedia retrieval and the evaluation of related systems. The particular challenge in this domain manifests in three key aspects: Firstly, multimedia retrieval comes with a diverse set of problems, spanning the straightforward task of locating one or several specific items in a dataset to acquiring aggregations or derivations thereof.

Secondly, multimedia retrieval evaluations can involve both human-in-the-loop (interactive) and automatic, machine-only (non-interactive) settings both for the retrieval process itself \cite{Nguyen2021:Interactive} and the verification of results that follows. The interactivity poses a particular challenge and introduces complexities in both stages.

And finally, the very motivation for research into efficient and effective multimedia retrieval is the sheer growth of multimedia data that could be observed over the past decades, both in volume and variety. The quantity and diversity of which must also be dealt with when conducting multimedia retrieval evaluations.

In this paper, we make three contributions towards a standardization of how multimedia retrieval evaluations are conducted. Firstly, we present a theoretical model formalizing the diverse aspects constituting multimedia evaluation settings in both current and future scenarios. Secondly, we map the concepts outlined in the model to practical applications in real-world evaluation campaigns. And finally, we introduce the open-source Distributed Retrieval Evaluation Server (DRES), which was first demonstrated in~\cite{DBLP:conf/mmm/RossettoGSBS21} and has since been used in multiple interactive multimedia retrieval evaluations~\cite{vitrivrVsSomhunter,DBLP:journals/access/TranNDHSLPNKDJRYAATHSG23,DBLP:journals/mms/LokocABDGMMNPRSSSKSVV23,DBLP:journals/ijmir/HellerGBG0LLMPR22,schall2023interactive,DBLP:conf/soict/DoNNTHGNLNNDT23}.

The remainder of this paper is structured as follows: \Cref{sec:relatework} provides an overview of the history of performance evaluation in multimedia retrieval and summarizes currently used methods.
\Cref{sec:model} introduces our formal model and \Cref{sec:applications} discusses its applications in real-world scenarios.
\Cref{sec:dres} then presents DRES,
which we offer as an open-source implementation of the proposed evaluation model.
Some practical applications of and use cases for DRES are then discussed in \Cref{sec:scenarios}.
Finally, \Cref{sec:conclusion} offers some outlook and concluding remarks.

\section{Performance Evaluation in Multimedia Retrieval}
\label{sec:relatework}

The field of \emph{information retrieval} and, as such, \emph{multimedia retrieval} is rooted in experimentation and evaluations. That is, evaluations accompanied research in this domain since its inception. The mechanism of evaluating information retrieval methods with a test collection was first introduced by \citet{cleverdonCRANFIELDTESTSINDEX1967} at the College of Aeronautics in Cranfield, UK, back in the 1960s.
This series of experiments was initially concerned with appropriate indexing languages for libraries.
By creating a reference collection, in which each document was associated with a certain relevance to a pre-defined information need, \citet{cleverdonCRANFIELDTESTSINDEX1967} essentially created what later would become be known as the \emph{Cranfield paradigm}~\cite{voorheesEvolutionCranfield2019}.
In these first experiments, verification was a very laborious endeavor since the relevance was based on manual judgments, but it was a breakthrough since the technique allowed for reproducibility.
Furthermore, the Cranfield experiments incorporated a concept that these days is referred to \emph{known-item search (KIS)}, meaning that the information need can be satisfied by an item that is known to be contained in the (test) collection.
Researchers could run their information retrieval method, e.g., a ranking algorithm, against the same test collection as others, thereby obtaining comparable results.
The results were based on metrics that were specifically invented during these experiments, \emph{precision} and \emph{recall}~\cite{keenMeasuresAveragingMethods1966}, which to this date are commonly used in retrieval evaluations. The two metrics have an inverse relationship, which is why both measures are required for a quality statement~\cite{baeza-yatesModernInformationRetrieval2011}.

However, precision and recall have been criticized for various reasons, among them that each measure individually has limited expressiveness, that total knowledge of the test collection is required for the calculation, and that neither reflects on the interactivity of modern-day search. The Cranfield paradigm also faced its share of criticism over the years, for instance, that test collections were too small compared to real-world data.
Yet, the Cranfield experiments remained the de facto standard for information retrieval evaluations for several decades.

In 1991, efforts were started at the US National Institute of Standards and Technology (NIST) to create a new, standard dataset in the context of the \emph{Text REtrieval Conference (TREC)}. The TREC initiative essentially built on the Cranfield methodology and made adaptions where necessary~\cite{harmanInformationRetrievalEvaluation2011}.
Based on a \emph{test collection} of documents, 
a set of \emph{tasks} (in TREC terminology: \emph{topic}) were formulated.
Participating research groups had to submit their ranked result list for each topic in one batch. Subsequently, a manual assessment of the relevance of each task produced the evaluation results.
Other such ``batched'' evaluation efforts centered around language retrieval were started shortly after TREC. Notably, the Asian counterpart to TREC, the \emph{National Institute of Informatics Test Collection for IR Systems (NTCIR)}, in 1999, has been held every 18 months since then. NTCIR introduced English-Japanese translation tasks~\cite{kando1999overview} and eventually the field of patent retrieval evaluations~\cite{kando2002overview}. Similarly structured to TREC, natural language processing (of Asian languages) has been another central part of NTCIR.
In 1998, a document collection from Switzerland that contained articles in three languages, German, French, and Italian, was published~\cite{sheridanBuildingLargeMultilingual1998}, which led to the inception of cross-language retrieval tasks (CLIR) in TREC.
Finally, CLIR tasks for European languages moved from TREC to its own initiative in 2000: The Cross-Language Evaluation Forum (CLEF, later Conference and Labs of the Evaluation Forum) was born in 2000, including English, French, German, and Italian~\cite{ferroInformationRetrievalEvaluation2019} content.

Several of the aforementioned evaluation campaigns started to include multimedia retrieval tasks over the years. However, the vast majority did not directly consider the interaction between the user and a retrieval system and stuck to the Cranfield paradigm and the batched submission mode. Ultimately, this is a criterion to divide the major multimedia retrieval evaluation efforts today into the categories \emph{non-interactive} and \emph{interactive}.

\subsection{Non-interactive Multimedia Retrieval Evaluation}

In the non-interactive multimedia retrieval domain, the following initiatives have been established: 

\begin{description}
    \item[TRECVid] \cite{Smeaton2006:Evaluation,Awad2021:TRECVid} is an annual workshop spun off from the Text Retrieval Conference (TREC) in 2003. In the years since, it hosted various tasks related to video retrieval, including Video Instance Search, Copy Detection, Known-Item Search, and Ad-hoc Video Search. For each task, the organizers provide a dataset as well as common metrics for assessing the quality of the results produced and submitted.
    
    \item[ImageCLEF] \cite{Ionescu_2023} is part of the Cross-Language Education and Function (CLEF) and aims to provide \textit{``an evaluation forum for the cross–language annotation and retrieval of images''}.\footnote{\url{https://www.imageclef.org}} Held annually since 2003, it offers various image retrieval-related tasks from multiple domains, including medical images or environmental photography.

    \item[MediaEval] is an annual multimedia benchmarking initiative\footnote{\url{https://multimediaeval.github.io/}} hosting a multitude of multimedia analysis tasks, such as image retrieval for news articles~\cite{DBLP:conf/mediaeval/KilleLOED21}, media memorability prediction~\cite{DBLP:conf/mediaeval/KiziltepeCDHFH021}, visual sentiment analysis~\cite{DBLP:conf/mediaeval/HassanARHCHA21}, or music emotion recognition~\cite{DBLP:conf/mediaeval/TovstoganBP21}. The initiative is quite broad, and not all tasks have clear retrieval aspects.
\end{description}

Apart from these larger campaigns, each with its various tracks and sub-tasks, there exists a multitude of other and more specialized evaluation venues containing retrieval components, which all fall in the non-interactive evaluation category~\cite{DBLP:journals/corr/abs-2008-00744,DBLP:conf/mm/CurtisAGS23}. While these are sufficient to evaluate certain aspects of retrieval methods in isolation, they  fail to reflect the end-to-end experience of somebody using a retrieval system in practice.

\subsection{Interactive Retrieval Evaluation}

Central to the Cranfield experiments' user model~---as this was a reasonable assumption these days---~were indexers or professional (re-)searchers: Human agents familiar with the domain and process of handling large knowledge collections.
Other than that, interaction with users who seek precise information through searching~---searchers~\cite{dattaImageRetrievalIdeas2008,voorheesEvolutionCranfield2019}---~has been very limited in the early days during the Cranfield experiments.
Nevertheless, \citeauthor{keenPerformanceNinePrinted1978} (\citeyear{keenPerformanceNinePrinted1978}), previously known for his work on metrics in the Cranfield series (cf~\cite{keenMeasuresAveragingMethods1966}) started work that focused more on the aspect of the searcher~\cite{michaelkeenABERYSTWYTHINDEXLANGUAGES1973,keenPerformanceNinePrinted1978}.
Major early-day work has been conducted by~\citet{belkin1980anomalous} and introduced the hypothesis that end users typically are not able to precisely formulate their information need, and as such, user queries could be grouped by requirements to the information retrieval system~\cite{belkin1980anomalous,belkinASKINFORMATIONRETRIEVAL1982}.
These early-day experiments would entirely focus on text retrieval since multimedia retrieval was not yet a dedicated field at the time. TREC incorporated human-in-the-loop processes to some extent ever since its inception, as the ranked result lists to be submitted could also be created manually~\cite{voorheesEvolutionCranfield2019}.
In general, these tasks did, however, not explicitly consider interactivity.
The first formalized \emph{interactive track} was introduced in TREC 3, aiming at comparing human and automatic routing efforts~\cite{overTRECInteractiveTrack2001}.
Interactive \emph{video} search roots in TRECvid's interactive video search track and opening up towards the field of interactive content-based search~\cite{snoekLearnedLexiconDrivenParadigm2007}.

While these evaluation initiatives provide essential and valuable results, they do not systematically consider the search activity itself as part of the evaluation. This paved the ground for the first interactive and \emph{competitive} evaluation efforts designed to~---apart from producing the evaluation results---~entertain a crowd (the conference participants) and demonstrate interactive (video) retrieval systems in use. A prime and early example of such a campaign was the VideOlympics~\cite{snoekVideOlympicsRealTimeEvaluation2008a}.
Built similarly to TRECvid ad-hoc search tasks, VideOlympics had its infrastructure specifically tailored to the interactivity of the event: an evaluation server displayed a scoreboard of the results compared to a ground truth in real time.
The Video Browser Showdown (VBS)~\cite{Schoeffmann2014:User,Lokovc2018:Influential,Schoeffmann2019:Video} to this day follows and expands upon this paradigm by providing tasks in three distinct categories;
\begin{enumerate*}[label=(\roman*),itemjoin={{, }}, itemjoin*={{, and }}, after={{}}]
\item Textual Known-Item Search (T-KIS)
\item Visual Known-Item Search (V-KIS)
\item Ad-hoc Video Search (AVS)
\end{enumerate*}~\cite{DBLP:conf/mmasia/SchoeffmannLB20}.
The first two task types build upon the very same principle of Cranfield's Known-Item Search~\cite{cleverdonCRANFIELDTESTSINDEX1967} and only differ in the modality used to present the information need.
For T-KIS, a textual representation of the information need is presented, and for V-KIS, the very portion of the video to be found is previewed. AVS tasks are structurally closer to TRECvid in that human judges assess submissions for a short (broad) textual information need. Notably, for AVS, a large and diverse set of results is expected.
In its latest installment, in 2024, VBS operated on three distinct data sets: One being very large and diverse in nature -- the Vimeo Creative Commons Collection (V3C)~\cite{rossetto2019v3c}, and the other two much smaller but highly homogeneous, sporting diving and underwater footage -- the Marine Video Kit (MVK)~\cite{DBLP:conf/mmm/TruongVHLTJY23} -- and videos of endoscopic surgeries. 

Competition-style benchmarking campaigns are also used by other multimedia retrieval communities, such as in lifelog retrieval \cite{Xu2021:Lifelog} with the Lifelog Search Challenge (LSC) \cite{Gurrin2022:Introduction,Gurrin2023:Introduction}.
Similarly to VBS, LSC used T-KIS tasks in addition to Ad-hoc Lifelog tasks in its latest installment. Most recently, Question and Answering tasks were also employed, in which an answer to a question had to be provided based on information found in the LSC dataset~\cite{lscdata:2019}. An example of such a question could be \textit{``What's the name of the restaurant I frequently visit when traveling to Oslo?''}. These types of analytics tasks differ in that the solution to the task is not part of the test collection but instead must be derived from its content, in this example, by finding the right items and reasoning about them. Therefore, formally, the solution is a specific derivative of items in the test collection.

In light of these different types of campaigns, some of the more recent work also aims to provide frameworks for interactive (multimedia) retrieval efforts~\cite{Lokoc2022:Task} by categorizing the space of tasks that were used in the past and might be used in the future. Additionally, these types of interactive settings also allow for the analysis of aspects that go beyond the mere correctness of the results that have been submitted. For example, \cite{Lokoc2019:Interactive} leveraged \emph{interaction logging} during the VBS 2018 installment to analyze what features the systems provided were used most frequently and successfully. This study was further built upon in  \cite{Lokoc2021:Is}, where an in-depth study was performed for the three top-performing systems that participated in VBS 2020.

\section{A Model for Performance Evaluation in Multimedia Retrieval}
\label{sec:model}
% Load model-related commands for easier editing
\input{model-commands.tex}
Irrespective of the specific domain of information retrieval evaluation, in its most general form, an evaluation can be described as consisting of the following six core components (loosely based on \cite{Smeaton2006:Evaluation} ``formula for TREC''):
\begin{description}
    \item[Test collection] is a collection of media items~---which we will henceforth refer to as \emph{documents}---~from which results are produced during an evaluation.
    \item[Tasks] describe an actionable search desire and information need as well as the constraints under which that information need must be fulfilled. Tasks constitute a major building block of every evaluation, which typically consists of a range of tasks that must be solved.
    \item[Agents] are the entities that solve a task. Typically, the agent is a tandem of the human operator and the instance of the retrieval system they interact with. However, other constellations are possible (e.g., machine-only).
    \item[Evaluation run] is a concrete instance of the evaluation with the agents, aiming to solve the tasks. The differentiation between an evaluation and a run implies that~---in principle---~an evaluation can be run multiple times.
    \item[Relevance judgment] is the process of assigning relevance (either binary or continuous) to the solutions proposed by the agents with regard to each task, e.g., the documents in the test collection.
    \item[Analysis] describes the analysis of the metrics recorded during the evaluation, which is particularly important in comparative studies.
\end{description}

Inspired by the three pillars, \emph{data}, \emph{task} and \emph{user} proposed by~\cite{datataskuser}, we introduce a separation of an evaluation into three phases: The preparation phase before an evaluation is being run, the phase during which the evaluation is executed and the phase after the evaluation has concluded, and results are being analyzed. The individual phases are introduced in the next sections.

\subsection{Evaluation Definition}

Before any evaluation can be executed, it has to be defined in such a way that the execution is actionable.
Therefore, we propose a formal model of such a definition in this section. While \cite{datataskuser} restricted the first pillar to the test collection an evaluation operates upon, we extend this idea also to include the definition of the tasks. In doing so, we acknowledge the importance of both the test data as well as the careful design of tasks and the associated description of the information need.

In classical, text-based information retrieval, oftentimes, the unit of retrieval is a \emph{document}~\cite{harmanInformationRetrievalEvaluation2011}, which is an element of the test collection.
However, particularly for \emph{dynamic multimedia data}~---that is, multimedia data that exhibits temporal progression such as videos~\cite{blankenMultimediaRetrieval2007}---~the unit of retrieval is often based at a sub-document level, for example, a temporal segment of a video. In some instances, the information need may not directly be included in the test collection at all but instead be some derivation of the data (e.g., a piece of information that can be extracted from an image). To accommodate both aspects, we distinguish between the \emph{document} $\mediaitem \in \collection$ as an element of the \emph{test collection} $\collection$ and the \emph{fragments} $\fragment$ the documents are comprised of. Our model is strictly agnostic regarding the semantics used to form these fragments. Consequently, a fragment can represent any part or derivation of the content of a document, such as a shot in a video, a region in an image, or a piece of information contained in the content. We denote the \emph{fragment set} $\fragmentset_\collection$ as the (potentially infinite) set of all possible fragments $\fragment \in \fragmentset_\collection$, given some test collection $\collection$.

Central to an evaluation following the Cranfield paradigm is a notion for information need descriptions, also referred to as \emph{search need presentation}~\cite{DBLP:journals/mms/LokocABDGMMNPRSSSKSVV23} and corresponding relevance judgments. As \cite{DBLP:journals/mms/LokocABDGMMNPRSSSKSVV23} points out, such search need presentations may themselves exhibit a temporal progression and can thus change over time. We, therefore, formalize the \emph{task description} \desc{} as a class of functions that map some notion of time \timeset{} to a set of fragments. 
The function's co-domain $\powerseta{\fragmentset}$ includes the empty set in case no task description exists at a given point in time.
Essentially, the \desc{} function presents such search need presentations in the form of \emph{fragments} for a given point in time or none. A concrete example of such a \desc{} function is described and visually represented in~\Cref{sec:task-desc}.

\begin{equation}
\label{eq:desc}
    \desc\colon\timeset\rightarrow\powerseta{\fragmentset}
\end{equation}

Relevance judgments, sometimes also referred to as the ground truth, indicate which fragments are relevant with respect to an information need description.
This is modeled by the \emph{relevance judgment} function \reljudge, which maps an answer set $\answerset$ to a verdict regarding its relevance. The notion of an answer set is properly defined in \Cref{eq:answerset}. For now, it can be thought of as a set of fragments an agent considered relevant and thus submitted as an answer for a task.
In order to accommodate binary and graded relevance, the semantics of the function's co-domain are as follows: No relevance is indicated by $0$, any form of (potentially gradual) relevance by $\left(0,1\right]$. The symbol $\relundet$ denotes \emph{undecidable} relevance, which can be useful in some corner cases.

\begin{equation}
    \label{eq:reljudge}
    \reljudge\colon \answerset \rightarrow \closedIntervala{0}{1}\cup\set{\relundet}
\end{equation} 

Collectively, the two functions $\desc$ and $\reljudge$ form a \emph{task template} $\tasktemplate$, which is a tuple as defined by \Cref{eq:tasktemplate}. It defines the two fundamental aspects of every task: how the information need is described to the agents and how potential solutions are evaluated against the ground truth.

\begin{equation}
    \label{eq:tasktemplate}
    \tasktemplate:=\tuple{\desc,\reljudge}
\end{equation}

In addition to these two required elements, a \emph{task template} can also encode arbitrary boundary conditions, such as a pre-defined running duration, as additional fields. A collection of $N$ \emph{task templates} can then be combined into an \emph{evaluation template} $\evaltemplate$, which outlines the content of an evaluation.

\begin{equation}
    \label{eq:evaltemplate}
    \evaltemplate := \lbrace \tasktemplate{}_1, \tasktemplate{}_2, \ldots , \tasktemplate{}_N \rbrace
\end{equation}

This evaluation template forms the foundation of the next phase -- the evaluation's execution.

\subsection{Evaluation Execution}
\label{sec:execution}

The evaluation definition phase is followed by its execution (the evaluation run). This phase follows the idea of creating \emph{task instances} $\task$ as elements in the set of tasks $\taskset$ based on the predefined \emph{evaluation template} $\evaltemplate$ and the \emph{task templates} $\tasktemplate$ it contains, leading to the minimal definition of an \emph{evaluation} \eval in \Cref{eq:eval}.

\begin{equation}
    \label{eq:eval}
    \eval := \tuple{\evaltemplate, \agentset, \taskset}
\end{equation}

Upon creation, the evaluation $\eval$ contains the evaluation template it has been created from as well as a set of participating agents $\agentset$. The task set $\taskset$ is empty at the beginning and populated with tasks $t$ as the evaluation progresses. Every task is an instance of a task template and can be regarded as a copy thereof, with extensions required to encode all the information accumulated during an evaluation.\footnote{Separating between the definition and the instance has several advantages. Most importantly, it allows for re-using the same information need description in multiple task instances, which may be a practical necessity.}

As tasks are being executed, searchable assignments described by the task $\task$ are acted upon by the participating agents $\agent \in \agentset$. Based on the description generated by a task's $\desc$ function, these agents aim to satisfy the information need within the constraints imposed and submit their findings, commonly in the form of a ranked list, which are then evaluated by a task's $\reljudge$ function. 

We model these result lists as \emph{answer sets} $\answerset$, which are ordered lists of \emph{answer tuples} $\answer$ that contain a fragment (the proposed solution) and an optional parameter $\answerscore$.

\begin{equation}
    \label{eq:answerset}
    \answerset := \lbrace a = \tuple{ \fragment, \answerscore } | \fragment \in \fragmentset \wedge\ \answerscore \in [0, 1] \rbrace
\end{equation}

Using \Cref{eq:answerset}, the \emph{submission} $s$ provided by an agent $\agent$ can be defined as the triple outlined in \Cref{eq:submission}, with $\submissiontime$ representing the point in time relative to the start of the task the agent submitted the answer set.

\begin{equation}
    \label{eq:submission}
    \submission := \tuple{\answerset,\agent,\submissiontime}
\end{equation}

This idea of a submission leads us to the structure of a task $\task$ in an evaluation. The task references the task template $\tasktemplate$ it has been derived from. To hold all submissions encountered in the course of its execution, the task must also hold a set of all submissions $\submissionset$.

\begin{equation}
    \label{eq:task}
    \task:=\tuple{\tasktemplate,\submissionset}
\end{equation}

As with the templates, both the task $\task$ and the evaluation $\eval$ may exhibit additional attributes depending on the concrete use case. Equations \ref{eq:eval} and \ref{eq:task} therefore constitute the minimal, viable definitions, which may be tailored to a particular application. For example, both evaluations and tasks typically have a defined start and end time in practice, which can be included as fields. 

\begin{figure}
    \centering
    \input{figures/run-state-machine}
    \caption{State-machine representation of the evaluation execution phase. The preceding (preparation) and following (analysis) have been omitted for the sake of brevity.}
    \label{fig:run-state-machine}
    \Description{A state-machine representation of the evaluation execution phase.}
\end{figure}
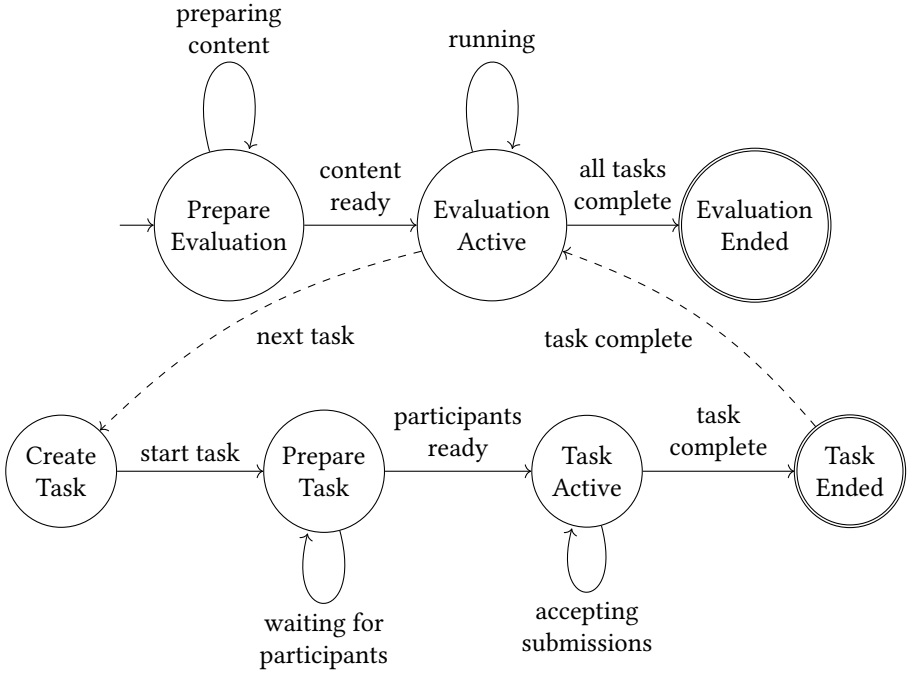

The basic evaluation execution process is illustrated in \Cref{fig:run-state-machine} as a pseudo finite automaton. Generally speaking, once an evaluation is started (after its initial definition), an evaluation goes into a \emph{preparation} state, which may include the distribution of the test collection or other preparatory steps, such as pre-processing of task descriptions. Once prepared, the evaluation becomes active and enters the \emph{running} state, during which tasks are being created from a template, prepared, and executed.

Every time a task is activated, it undergoes a creation and preparation stage. Once preparation of a task has concluded, it enters the \emph{running} state, during which submissions by the agents are accepted. Once a task has finished, it transitions into the \emph{ended} stage, and the next task may begin. Once all tasks have concluded, the evaluation ends. A task's transition from \emph{running} to \emph{ended} is task-specific. For example, a particular task may be considered solved as soon as every participating agent has submitted at least one answer. Alternatively, a task may have a predefined runtime and may end as soon as the time has elapsed.  This is, in fact, true for most transitions, which may be either driven by some internal logic of the system or by a human conductor of the evaluation. 

With respect to how evaluations are conducted, and based upon the classification of evaluations presented in \Cref{sec:relatework}, one can (roughly) distinguish between the following three modes:

\begin{description}
    \item[Interactive-synchronous Evaluation] At any point in time, every participating agent $\agent$ is engaged in at most one task $\task$ (interactive), which must be identical for all participating agents, including synchronized start and end time (synchronous).
    Interactive-synchronous evaluations often have a human conductor orchestrating the evaluation. Typically, tasks have a short run time, and collectively, they constitute a progression over the course of an evaluation, which is the same for all agents. 
    \item[Interactive-asynchronous  Evaluation] At any moment in time, every participating agent is engaged in at most one task $\task$ (interactive), which may, however, differ between different agents (asynchronous).
    For such evaluations, the logic of how the evaluation progresses is typically automated. In practice, the order of tasks might be different for every participating agent. Tasks typically also have a short time span.
    \item[Non-interactive Evaluation] At any moment in time, any participating agent can be engaged in multiple tasks (non-interactive). Non-interactive tasks are inherently asynchronous. Typically, such tasks are long-running and can be solved in any particular order. Usually, no human orchestration is involved (aside from the initial definition). These task types resemble the early batched submission evaluations.
\end{description}

\subsection{Evaluation Analysis}
\label{sec:analysis}

The third and final phase involves the evaluation analysis. It may take place in real-time during the evaluation as well as after the evaluation execution has concluded. In analogy to \citeauthor{datataskuser}'s \cite{datataskuser} third pillar (the user), the evaluation analysis consists of ``judging the success or failure of the similarity-based application [the MR system]''~\cite[p. 327]{datataskuser}. In this sense, all the submissions made are analyzed, and conclusions are drawn from the analysis during this phase.

The formal definition of all entities involved in evaluation definition and execution forms a solid foundation for such an analysis. For a multimedia retrieval evaluation, arguably, the \emph{submissions} introduced in \Cref{eq:submission} are one of the core entities to work with, especially when answering questions about the effectiveness of individual agents.  At a high level, we identify two types of metrics that can be derived directly from the entities described. \emph{Task metrics}, as defined in \Cref{eq:taskmetric}, take an agent $\agent$ and a task $\task$ as input and outputs a non-negative, real-valued score that indicates how well the agent $\agent$ performed in the task (higher is better).

\begin{equation}
    \label{eq:taskmetric}
    \taskmetric\colon \agentset \times \taskset \rightarrow \realsetpos
\end{equation}

Aggregations of such task metrics are possible in any form. However, from the perspective of an evaluation, it is highly likely to aggregate over the entire evaluation.
Therefore, we propose the class of \emph{evaluation metrics} analog to task metrics, defined per participating agent $\agent\in\agentset$ and evaluation $\eval\in\evalset$ in contrast to per task:

\begin{equation}
    \evalmetric\colon \agentset \times \evalset \rightarrow \realsetpos
\end{equation}

Canonically, an evaluation metric \evalmetric{} is expanded to the tasks of an evaluation \eval.
For both types of functions, the domain can be extended by an arbitrary number of auxiliary parameters.  The types of analyses can, however, be vastly extended by extending the basic data model described thus far. For example, submissions $\submission$ can also be extended to contain information about user-system interaction. Such interaction logging was shown to be highly valuable when it comes to distinguishing between the performance of a retrieval system as opposed to its human operator.

\section{Practical Applications of the Model}
\label{sec:applications}

The model for performance evaluation described in \Cref{sec:model} provides a formal framework for multimedia retrieval evaluations. In this section, we revisit parts of that model and approach it from a more practical point of view in order to further illustrate its application in multimedia retrieval evaluation.

\subsection{Tasks and Information Need Description}
\label{sec:task-desc}
With regard to the task description function we have presented in~\Cref{eq:desc}, we acknowledge that in a multimedial practice, this formalization is often simplified considerably.
The semantics of said function is to describe the relevant information need to a participating agent.
We model this mechanism of conveying the relevant information as a set of temporally aligned \emph{channels} with one channel per information type or modality.
For every channel, only a single fragment can be presented at a given point in time.
While one could imagine arbitrarily many different types of channels conveying different information, in practice, the information is restricted to what can easily be presented using common audio-visual output devices.
Commonly used channels include text, image, video, and audio.

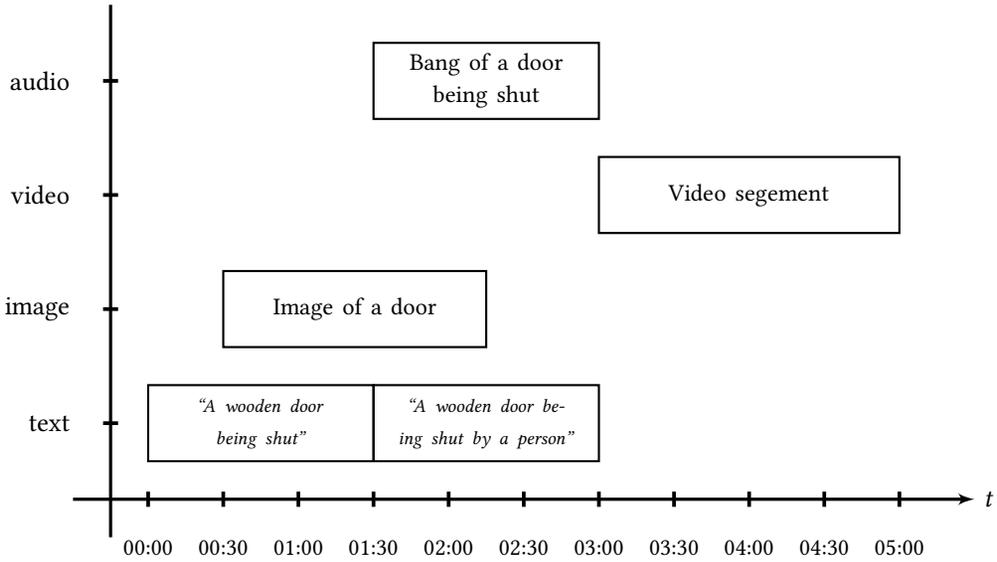
\begin{figure}
    \centering
    \input{figures/query-hints.tex}
    \caption{Illustration of an information need description along different media channels. Rectangles indicate the usage of a channel with a given start and end time. The concept of media channels for task descriptions limits the powerset returned by the task description function described in \Cref{eq:desc}.}
    \label{fig:query-hints}
    \Description{An illustration of an information need description along different media channels. Rectangles indicate the usage of a channel with a given start and end time. The concept of media channels for task descriptions limits the powerset returned by the task description function described in Equation 1.}
\end{figure}

To illustrate this, we present an example of a task with a multi-modal information need description in \Cref{fig:query-hints}. This example task aims to find a video segment in the test collection. The task's description starts with text (first channel), to which, after 30 seconds, an image is added (second channel). At this point in time, agents can leverage both the textual description and the example image to solve the task.  After 90 seconds, the text is expanded, and the sound of a door being shut is played in a loop (third channel), which also adds the aural modality. In an interactive setting, such an aural cue could be used by the agent to verify items in the result set. Starting from the three-minute mark, the desired scene could be played as a video (fourth channel), transforming the task from a textual to a visual search task.

It is also to be noted that while the task description may vary from task to task, the overall form of that description is often shared among a set of tasks within an evaluation. For example, in V-KIS tasks, the task description is always a document from the test collection (or parts thereof). To simplify the definition process and implementation aspects, it may, therefore, make sense to introduce an additional abstraction we call \emph{task type}, that captures such common properties shared by multiple templates.

\subsection{Assessment of Relevance}

While the relevance function as defined by \Cref{eq:reljudge} makes no restrictions on how the relevance of a provided answer is determined, practical instantiations can be categorized along two independent axes, which we will discuss in the following.

Once again, and similar to the task's information need description, the type of relevance assessment is often shared by a set of tasks. Therefore, a similar argument can be made for the introduction of an additional layer of abstraction in the form of task types.

\subsubsection{Time of Determinacy}

The first distinction in the space of relevance functions is whether it is possible for a given task to enumerate possibly relevant answers \emph{a priori} or if it is necessary to render a verdict on a received answer \emph{a posteriori}. 

\emph{A priori} relevance assessment requires that for a given query and test collection, a complete annotation of relevance or an otherwise finite set of possible relevant answers exists. 
This is generally easy for certain types of tasks, such as known-item search scenarios, since only one fragment in the test collection matches the query, assuming the test collection is free of duplicates.
For sufficiently small test collections, it might also be feasible to densely annotate the entire collection with respect to the relevance to a particular query beforehand.

\emph{A posteriory} relevance assessment, in contrast, is applicable in all cases where it is not possible or feasible to assess any possible fragment of a test collection or where there might be an arbitrarily large set of relevant answers.
Such assessments are generally difficult to mechanize and rather rely on human assessors.
An example of a task type that requires this form of assessment is ad-hoc search, where a low-specificity query is given, and it is commonly not feasible nor desirable to densely assess the relevance of the entire test collection.

\subsubsection{Type of Answers}

The second distinction among relevance functions concerns their input.
Depending on the task type, the provided answer can consist of \emph{existing} fragments that are part of the test collection or \emph{derived} fragments that do not necessarily exist before a task is evaluated but are rather created during the evaluation.

\emph{Existing} fragments are commonly used in classical retrieval tasks, where a large amount of content needs to be reduced to a subset relevant to a given information need.
The answers for such tasks can be whole documents from a test collection or any part of such documents.

\emph{Derived} fragments, in contrast, are relevant in more analytical tasks, where the answer to a query can not be found in a test collection directly but has to be generated based on relevant information.
An example of this answer type is \emph{question answering} tasks, where a question's answer must be given in natural language.

\subsection{Evaluation Execution}

The lifetime of an evaluation run, as depicted in \Cref{fig:run-state-machine}, does not further discuss the events triggering any of its state transitions in order to be as generally applicable as possible.
In practice, however, there are a few statements that can be made that apply to commonly used scenarios.

A first and rather trivial statement is that the `next task' transition is generally triggered manually.
This is true for both synchronous and asynchronous interactive evaluations, although the actor who triggers the transition is not the same in both cases.
For synchronous evaluations, this transition is triggered by the evaluation conductor, while in the asynchronous case, each agent triggers the transition individually.
This distinction is also relevant for the following state transitions since the preparation and synchronization step is only explicitly relevant for synchronous cases, where it is important to ensure that all agents can start a task at the same time. In an asynchronous case, each agent's readiness can be implicitly assumed as soon as the `task start' transition happens.
A more in-depth discussion of the differences in state transitions between synchronous and asynchronous evaluations can be found in \cite{10.1145/3552467.3554797}. For non-interactive evaluations, the `next task' transition is also triggered by an evaluation conductor, although its semantics are slightly different. Since, in the non-interactive case, all tasks of an evaluation can be active in parallel, this transition happens simultaneously for all tasks rather than one task at a time.

Other broadly applicable statements regarding the `task complete' transition can be made. For practical reasons, there are several events that can trigger this transition. A commonly used condition is that of a maximum task duration. For interactive evaluations, this duration is commonly defined as the maximum time during which agents can attempt to solve a task (which is part of the task's definition). Non-interactive evaluations might rather have a global submission deadline, defining when answers for any task are accepted. Depending on the type of task and the form of expected answers, there might be additional triggers to end a task before its maximum time is exhausted.
This can happen if, for example, all participating agents have already solved the task by providing an answer that is evaluated to be correct. Similarly, a task might only accept a fixed maximum of answers, independently of their correctness, i.e., every agent only gets a single attempt to solve the tasks. For synchronous interactive evaluations, these conditions need to be fulfilled by all agents in order to trigger the premature end of a task. In asynchronous cases, each agent can trigger them individually.

\subsection{Evaluation Analysis}

In order to quantify the performance of any agent, a set of functions is required that represents this performance numerically, as outlined in \Cref{sec:analysis}.
Commonly used quantification functions include precision, recall, reciprocal rank, discounted cumulative gain, etc. These measures are commonly defined such that a higher number represents a `better' performance. They are hence also referred to as \emph{scoring functions}.

Depending on the task type, more elaborate scoring functions might be used that take into account multiple aspects of provided answers concurrently. For example, for interactive evaluations, it might not only be interesting to evaluate if a provided answer is correct but also how long it took an agent to provide such an answer and/or how many attempts the agent needed before submitting the correct one, penalizing delay as well as mistakes. Scoring functions taking these aspects into account are described in \cite{Lokovc2018:Influential}, namely for V-KIS, T-KIS, and AVS task types, and used during campaigns such as VBS or LSC.

In addition, and since an evaluation may contain different types of tasks that, in turn, make use of different scoring functions, it may also become necessary to be able to define a set of \emph{aggregation functions}. These are basically scoring functions that build a compound score out of individual scores. Often, these also include normalization aspects, as evaluations commonly compare the relative performance of agents rather than assigning absolute values.

For interactive evaluations, especially synchronous ones, these scoring and aggregation functions are commonly evaluated in real time. This not only has the benefit of providing a continuous readout of the current evaluation state, but it can also be used as a scoreboard showing the current ranking of all agents for information and entertainment. Again, this is common practice in campaigns such as VBS or LSC.

\section{The Distributed Retrieval Evaluation Server}
\label{sec:dres}
The theoretical model outlined in \Cref{sec:model} and the practical aspects described in \Cref{sec:applications}, form the foundation for the \emph{Distributed Retrieval Evaluation Server} (DRES)~\cite{DBLP:conf/mmm/RossettoGSBS21}.
The implementation of DRES is freely available as open-source software\footnote{\url{https://dres.dev}} and has already been used for several larger-scale evaluations. Most notably, it has served as the base infrastructure for the annual Lifelog Search Challenge (LSC) \cite{Gurrin2022:Introduction,Gurrin2023:Introduction} since 2020 and for the Video Browser Showdown (VBS) \cite{Lokovc2018:Influential,Schoeffmann2019:Video} since 2021. This section provides an overview of DRES' architecture and relevant implementation details.

\begin{figure}[tb]
    \centering
    \includegraphics[width=\textwidth]{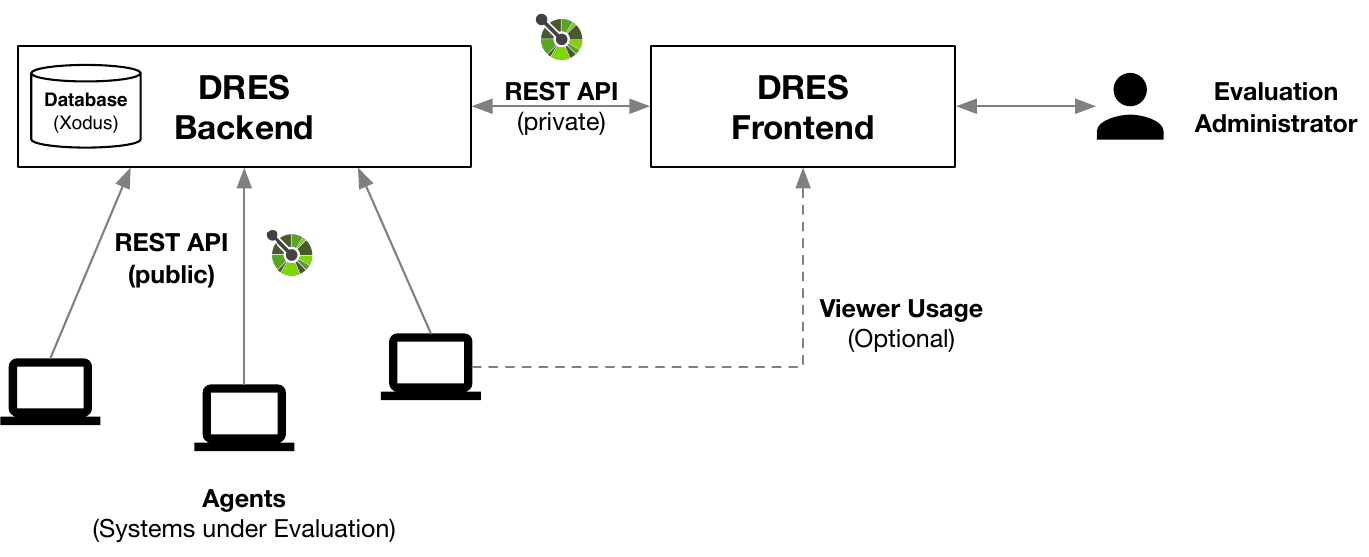}
    \caption{Overview of DRES' architecture and system components, which include the backend and frontend as well as a persistence layer.}
    \label{fig:dres-architecture}
    \Description{An overview of DRES' architecture and system components, including the backend and frontend and a persistence layer.}
\end{figure}

\subsection{Architecture}

DRES is designed as a web application accessible via a browser and it provides all the relevant functionality via a set of REST APIs. This setup was chosen to minimize the requirements on the side of evaluation participants (i.e., the agents) and to facilitate the remote evaluation of retrieval systems in both synchronous and asynchronous settings.

An architecture overview of DRES is provided in \Cref{fig:dres-architecture}. DRES consists of two components: a backend application written in Kotlin, which handles the evaluation state and provides API access to all relevant functionality, and a frontend component written in TypeScript and built using the Angular\footnote{\url{https://angular.io/}} framework, which serves as the primary channel for all user interaction. The state of every evaluation is persisted in a database (JetBrains Xodus\footnote{\url{https://github.com/JetBrains/xodus}}).

The browser-based user interface can be used for setup and administration, e.g., by an evaluation organizer. In interactive settings, it is also the main channel for communicating task descriptions to agents and informing participants about the overall progress of an evaluation and its tasks. This also includes features geared towards competitive live settings, such as live score updates.

Agents submit their answers via a public REST API, for which we provide a machine-readable OpenAPI specification. This specification can be used to automate the implementation of client-side stubs, which simplifies the interfacing process with the submission backend, regardless of the client's programming language. The backend- and frontend components also communicate via a (private) REST API, which leverages the same technology.

\subsection{Data Model}
\label{sec:datamodel}
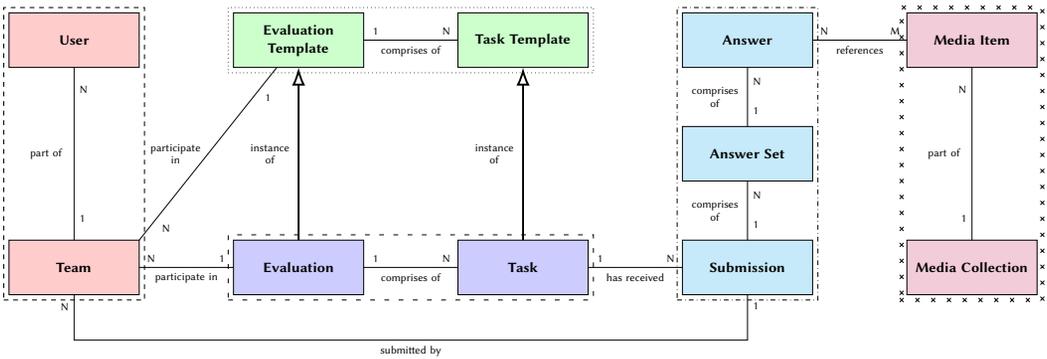
\begin{figure}[bt]
    \centering
    \resizebox{\textwidth}{!}{\input{figures/dres-erd.tikz}}
    \caption{Overview of DRES' data model. Only the most important entities are illustrated. Entities are grouped as follows: densely dashed group entities related to the agents, entities in the dotted group related to templating (evaluation definition), loosely dashed group entities related to the evaluation execution, dash and dot group entities used to model submissions, and the crossed groups entities representing the test collection.}
    \label{fig:dres-erd}
    \Description{An overview of DRES' data model. Only the most important entities are illustrated. Entities are grouped as follows: densely dashed group entities related to the agents, entities in the dotted group related to templating (evaluation definition), loosely dashed group entities related to the evaluation execution, dash and dot group entities used to model submissions, and the crossed groups entities representing the test collection.}
\end{figure}

The most important entities in DRES' data model are illustrated in \Cref{fig:dres-erd}. The green entities~---namely \emph{evaluation template} and \emph{task template}---~are directly related to the evaluation definition phase. These entities can be used to describe and define the evaluations and the tasks they consist of. This includes metadata, the definition of task description and relevance judgement functions and the information about a task's duration. As described in \Cref{sec:model}, these templates are then instantiated into a concrete implementation, which are highlighted in blue~---~namely \emph{evaluations} and \emph{tasks}. At this level, DRES records information about the start and end of a particular task instance.

The entities marked in red, i.e., \emph{team} and \emph{user}, represent the agents participating in an evaluation: A user is mapped directly to some person or system that can submit to a DRES instance and it is also the entity that is used for DRES' authentication and authorization sub-system. Users can be combined into teams if they are meant to cooperate in solving tasks, which affects how submissions are counted and tallied. In that regard, the team is what we consider the agent in the theoretical model, which can consist of one or multiple participating users. This supports the evaluation of collaborative retrieval approaches while still being able to distinguish submissions made by individual team members. 

Media collections, i.e., the test collections used for the evaluation, are also mapped to special entities. In fact, DRES offers an entire module dedicated to managing media collections and the items they contain.

\subsection{Submissions}

The entities related to \emph{submissions} are colored in cyan in \Cref{fig:dres-erd}. The data model is very much aligned with the model described in \Cref{sec:execution}.
That is, the submission-related entities directly align to the formal \submission, defined in~\Cref{eq:submission}, where the submission entity has a one-to-many relationship with the answer set entity, which in turn has a one-to-many relationship with the answer. 
Every submission is assigned to the submitting user and team and timestamped upon reception. In deviation from the model and for practical purposes, a single submission can comprise one or many \emph{answer sets}. This allows for optimizations, such as batching multiple answers into a single HTTP request. For this reason, it is also possible to specify the task ID or task name in an answer set in case sets aimed at different tasks are batched together. This is particularly useful for non-interactive evaluation settings. In line with the original definition, an answer set comprises of one to many \emph{answers}, which also allows for more complex scenarios in which not only a single but multiple answers are sought. An individual answer, in turn, can specify either an arbitrary text, a document, or a temporal segment of a document, which covers most but not all of the cases allowed by the definition of a fragment in \Cref{sec:model}.

\begin{lstlisting}[caption={Example of a submission, in which a temporal segment of a single video document that is part of the test collection is being submitted. Specifically, the submission describes video item ``v-09679'' between second 15 and 16.},captionpos=b,label={lst:submission},float,basicstyle=\footnotesize\ttfamily\color{black},frame=single]
{ 
  "answerSets": [
    {
      "answers": [
        {
          "mediaItemName": "v-09679",
          "start": 15000,
          "end": 16000
        }
      ]
    }
  ]
}
\end{lstlisting}

Individual submissions can be posted to a specified API endpoint, which requires authentication and is, therefore, only available to the users specified in a running evaluation. The data structure of a submission is depicted in Listing \ref{lst:submission} as a JSON.

\subsection{Synchronous and Asynchronous Evaluations}

Synchronous evaluations were the first use case for DRES. In the synchronous case, the evaluation administrator serves as a central conductor, determining the start (and end) time of a task, as well as the task order and the progression within an evaluation.
This is an intuitive approach in a localized evaluation setting, where all participants share a common location and, therefore, common experimental conditions.
In such cases, task information is commonly presented to all participants using a single instance of the user interface, for example, by projecting it onto a large screen.
But even in distributed settings, where each participant requires a separate instance of the user interface, DRES ensures that the state of these instances is synchronized, such that the same information is made available to all participants simultaneously.

In a setting where participants do not share a common location, synchronicity in tasks might be unnecessary or even undesirable, such as when participants are distributed across a larger number of time zones. Such distributed scenarios can benefit participants by enabling them to solve the tasks independently. Therefore, in addition to the \emph{synchronous} interactive evaluation scenario, in which all participants have to solve the same task simultaneously, DRES also supports an \emph{asynchronous} mechanism, where participants can be evaluated independently of each other.
This mechanism was first introduced in \cite{10.1145/3552467.3554797}, where it is described in more detail. Asynchronous evaluations are intended for remote settings and grant agents more control over their individual progression within an evaluation, independent of the other agents.

To facilitate this, DRES shows the participant a participant-specific view of the evaluation state, which excludes information about other participants' actions (other than the scores they have achieved). The evaluation administrator is still in charge of defining the tasks of an evaluation but no longer acts as a conductor/coordinator for the evaluation itself.
Instead, participants indicate to DRES their readiness for the next task and solve it within the (time-)constraint outlined in the definition. As soon as the task is solved or the time during which solutions are accepted is over, the option to start the next task becomes available.
This continues until all tasks of an evaluation have been presented or until the evaluation administrator closes the evaluation. This gives the agent full control over how the evaluation unfolds.

\subsection{User Interaction}

The DRES user interface is mainly geared towards supporting the first two phases of an evaluation life-cycle, namely, the evaluation definition and its execution. However, some basic analysis logic is also included, e.g., in the form of scoreboards that can be presented during a run. The underlying assumption is that the first two phases can be standardized using our proposed model, whereas the analysis is very individual to a particular evaluation setup. Nevertheless, the various data export capabilities and interfaces also offer some support in that regard.

\subsubsection{Evaluation Definition}
\begin{figure}[tb]
    \centering

    \begin{subfigure}{\textwidth}
        \includegraphics[width=\textwidth]{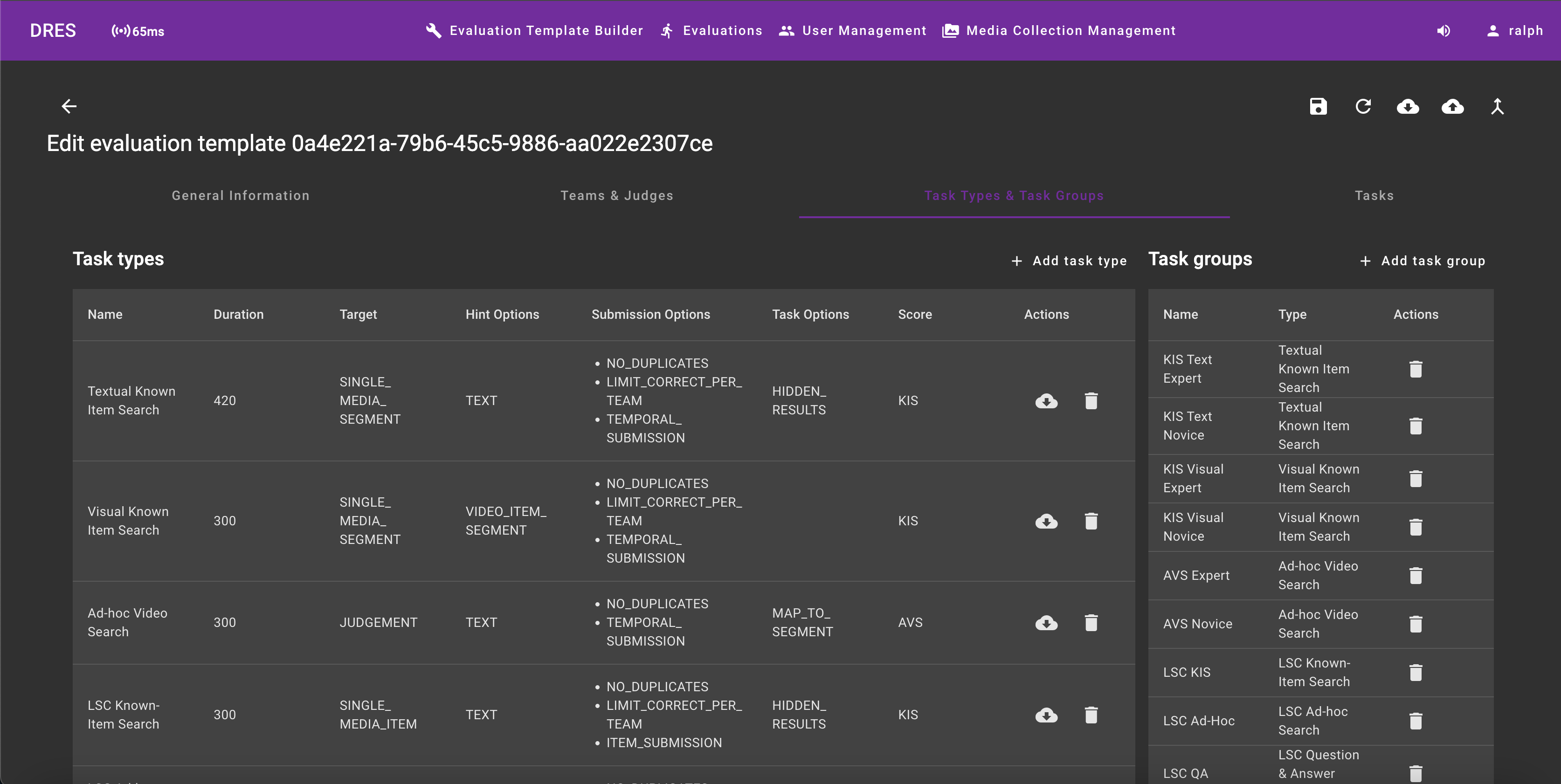}        
        \caption{The evaluation editor allows for changes to various aspects of an evaluation template, including the agents participating in the evaluation (teams), the types of tasks, and the individual task templates.}
        \label{fig:ui-definition-01}
    \end{subfigure}
\par\bigskip
\par\bigskip
    \begin{subfigure}{\textwidth}
        \centering
        \includegraphics[width=\textwidth]{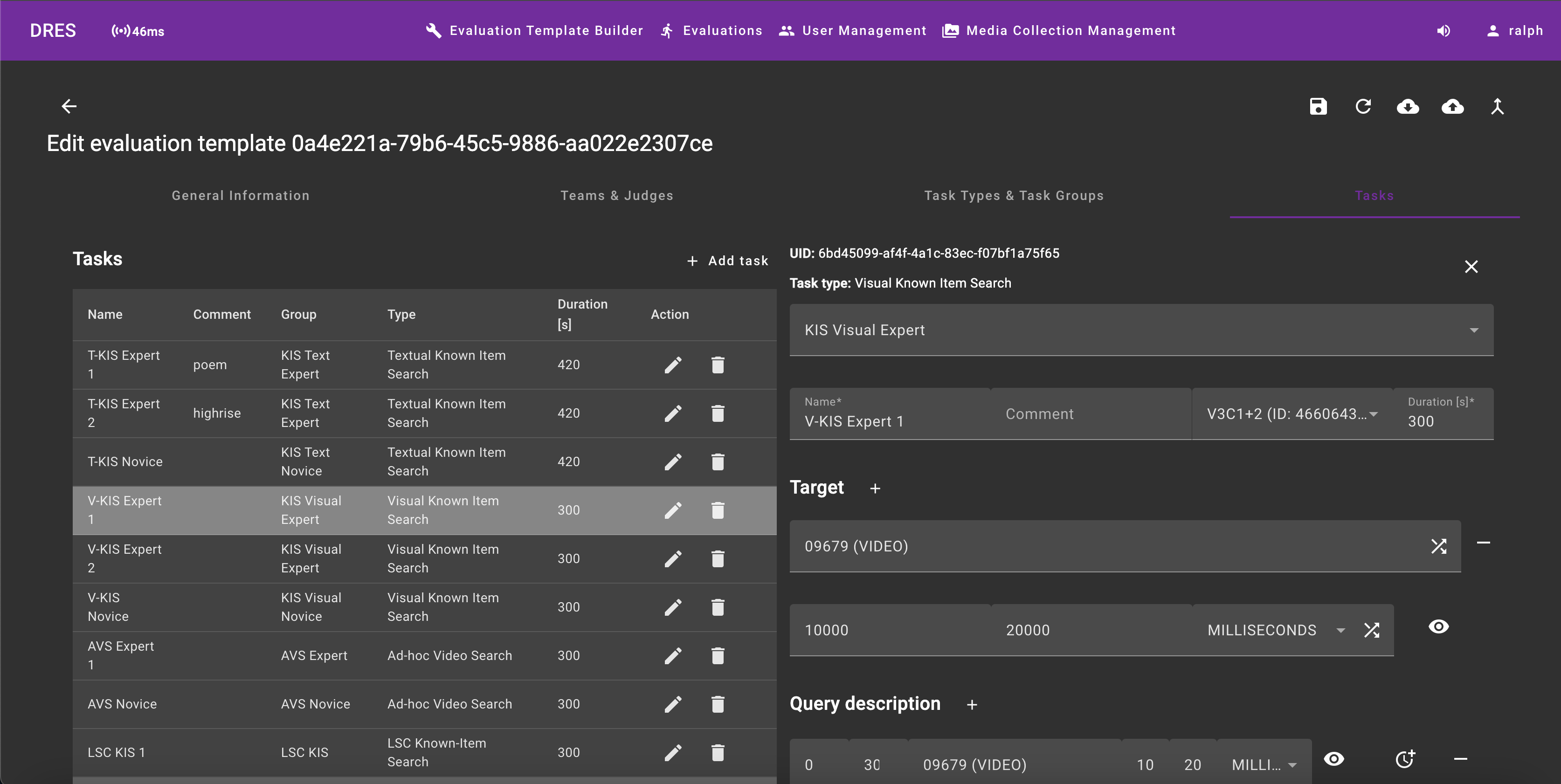}
        \caption{The task editor allows the authoring of individual tasks. The provided examples show a visual KIS task used during the Video Browser Showdown. The relevance judgment is provided by a known ground truth (the target). The task description is a segment of the video in question.}
        \label{fig:ui-definition-02}
    \end{subfigure}

    \caption{Screenshots of DRES' evaluation editor, which facilitates the management of evaluation templates.}
    \label{fig:ui-definition}
    \Description{Screenshots of DRES' evaluation editor, which facilitates the management of evaluation templates.}
\end{figure}

DRES' user interface provides a means for evaluation administrators and organizers to manage different evaluation templates and the task templates they contain. The UI for this -- the evaluation editor -- is depicted in \Cref{fig:ui-definition}. In addition to the meta information that describes the evaluation, the DRES data model supports the definition of the following aspects.

Firstly, an administrator can specify the participating \emph{teams}, which can be made up of different \emph{users} and which can be grouped into \emph{team groups}. The first two entities are described in \Cref{sec:datamodel}. Team groups simply provide an additional level of aggregation, mainly geared towards the later analysis stage. In addition to providing this evaluation-specific functionality, users are also the entity for which authentication and authorization are handled. It goes without saying that DRES also offers a user management module for this particular purpose.

Secondly, an administrator can specify \emph{task types} and \emph{task groups}. On the one hand, the task types define attributes shared among multiple tasks as described in \Cref{sec:applications}. This is mainly for convenience since every task template inherits said attributes from its type. On the other hand, task groups are an organizational feature. One use case during VBS or LSC is to distinguish between the tasks for the main session and those for the novice session of the evaluation.

Lastly, an evaluation organizer can author different task templates using the task editor. This editor's user interface is depicted in \Cref{fig:ui-definition-02}. Depending on the type of task description(s) specified in the task type, the editor can be used to select target items or segments that should be presented to the agents as hints when solving the task. This is complemented by a simple media collection module, in which documents (e.g., images or videos) and temporal segments thereof can be managed. That same module also provides the foundation for specifying the ground truth for relevance judgment. Alternatively, one can also specify textual descriptions or select external resources that can act as hints as well.

\subsubsection{Evaluation Execution}
Once an evaluation has been specified, the DRES user interface can be used to spawn any number of instances for a particular template. Once such an instance has been created, there are two user interface components for the evaluation execution.

\begin{figure}[tb]
    \centering

    \begin{subfigure}{\textwidth}
        \includegraphics[width=\textwidth]{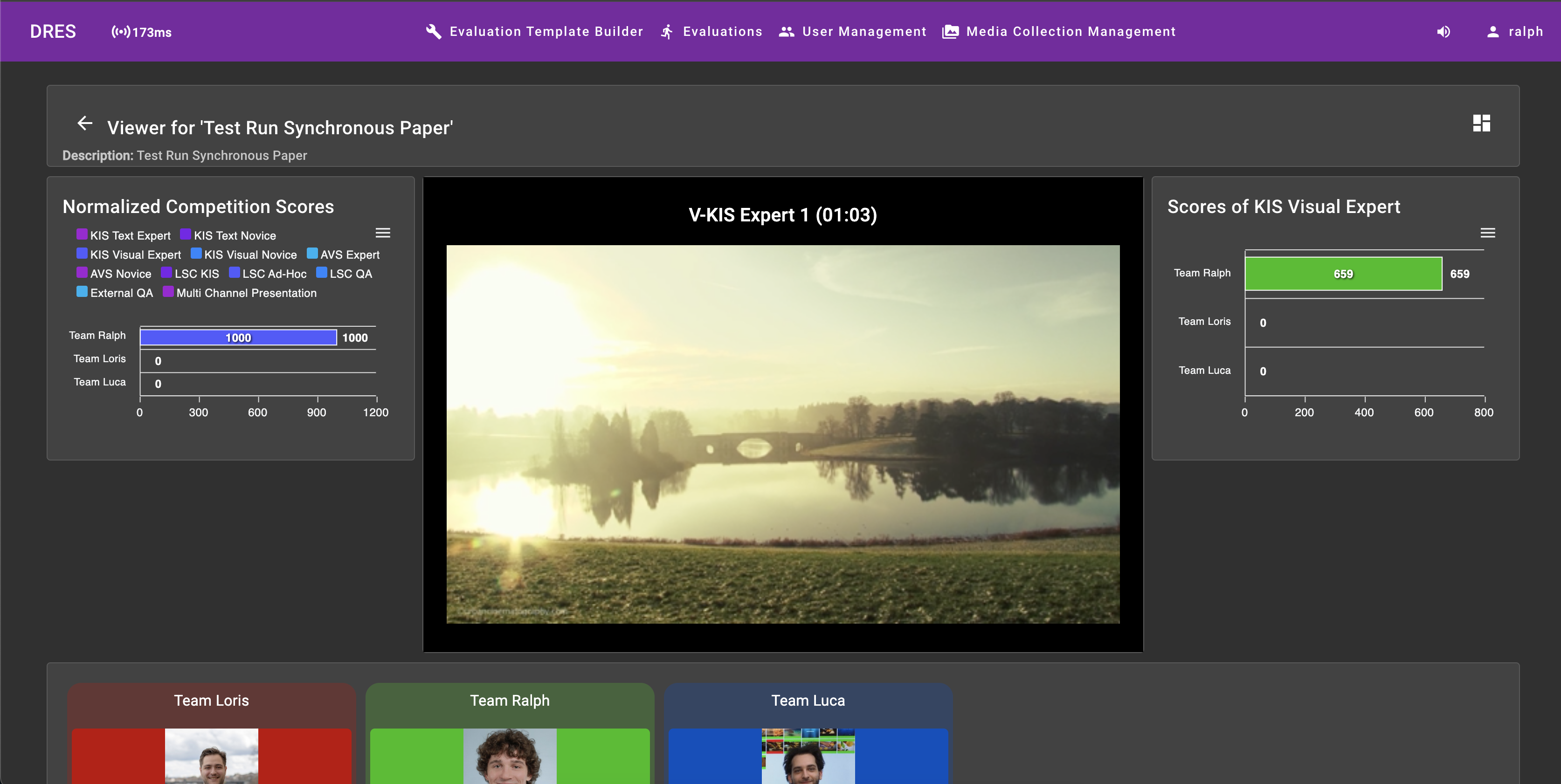}        
        \caption{The viewer is used to communicate the current state of the evaluation and the active task to all participating agents. This includes information about scores and, most importantly, the active task and its description. In this example, a video is presented (V-KIS). Also, the current scores for all the teams are being tallied and displayed.}
        \label{fig:ui-execution-01}
    \end{subfigure}
    \par\bigskip
    \par\bigskip
    \begin{subfigure}{\textwidth}
        \centering
        \includegraphics[width=\textwidth]{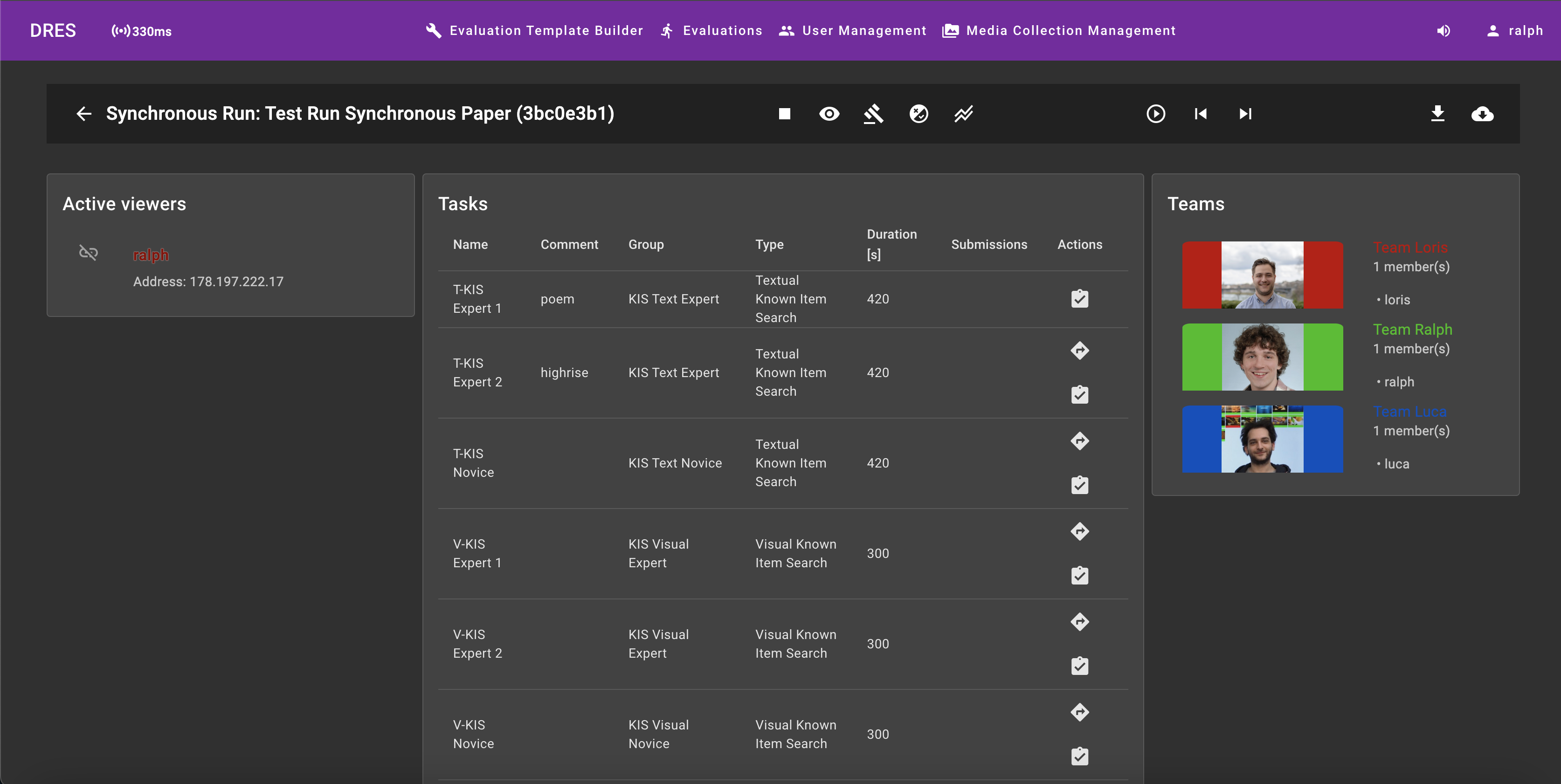}
        \caption{The task administration view provides the evaluation organizer the tools to moderate and orchestrate an evaluation. This includes basic functionality such as starting and ending an evaluation but also the ability to control the flow of an evaluation at a task level in synchronous settings.}
        \label{fig:ui-execution-02}
    \end{subfigure}

    \caption{Screenshots of DRES' task viewer and task administration view, geared toward evaluation execution.}
    \label{fig:ui-administration}
    \Description{A screenshot of the administrator user interface of DRES.}
\end{figure}

On the one hand, the viewer, which is depicted in \Cref{fig:ui-execution-01}, is used to communicate the evaluation state to the participating agents. This includes a summary of the scores attained by all the teams. Most importantly, however, the viewer can be used to display the currently active task, the time left for solving it, and the task descriptions that are currently being presented. To ensure fairness in a distributed, synchronous evaluation setting, a synchronization mechanism is implemented so that all the teams are presented with the specified descriptions simultaneously.

The evaluation administration, on the other hand~---which is depicted in \Cref{fig:ui-execution-02}---~gives the evaluation organizer the tools to moderate and orchestrate the evaluation. While useful for all evaluations, it is most useful for synchronous evaluation types. In this setting, the administrator can use the view to switch between tasks, start and end them (see \Cref{fig:run-state-machine}), or to increase or decrease the duration of running tasks, should the situation require it. Furthermore, that view can be used to adjust and overrule relevance judgments in case a mistake becomes apparent during the judgment process. All changes made to an evaluation are audited to ensure traceability and reproducibility of results.

\section{Application scenarios}
\label{sec:scenarios}

This section starts with providing an overview of evaluation experiments using DRES and continues by presenting several application scenarios demonstrating the practical applicability and usefulness of the introduced model and its implementation in DRES.

\subsection{Overview of DRES Usage}

Our implementation of the model, DRES, has been used in both, recurring evaluation benchmarks such as VBS \cite{DBLP:journals/ijmir/HellerGBG0LLMPR22,DBLP:journals/mms/LokocABDGMMNPRSSSKSVV23} or LSC \cite{Gurrin2022:Introduction,Gurrin2023:Introduction}, as well as 
other challenges, such as the Ho Chi Minh City AI Challenge 2023~\cite{DBLP:conf/soict/DoNNTHGNLNNDT23}.
Outside of formal challenge settings, DRES also found several applications in individual experiments~\cite{vitrivrVsSomhunter,willis2024task} and teaching and outreach activities at the Universities of Zurich, Basel, and Charles University in Prague.

Since VBS and LSC are annual evaluation campaigns, we present an overview of the past installments in~\Cref{tab:overview}.
We report on the number of participating teams,\footnote{In VBS'24, a team consisted of one single system operator. However, multiple teams could use the same retrieval system.}, the number of individual tasks solved, the number of categories of tasks, as well as the number of individual submissions during the evaluations.
For VBS, the reduction of submissions over the last years was by design.
Equally so for LSC, the increase in submissions is linked to more categories.
For LSC'23 and VBS'24, a new category requiring textual answers not part of the media collection has been introduced.

\begin{table}[ht!]
\def\arraystretch{1.25}
\setlength\tabcolsep{6pt} % 4pt seems default, 8pt is too much
\caption{An overview of the past installments of the Video Browser Showdown (VBS) and Lifelog Search Challenge (LSC) using DRES that have been using DRES. For VBS, up until 2024, teams incorporated multiple individual users, whereas, since 2024, there has been a single-user-team policy established.}
\label{tab:overview}

\resizebox{\textwidth}{!}{
\begin{tabular}{l|rrrr|rrrrr}%this is on purpose not tabularx with X, as I think it doesn't look as nice
\hline
            & \multicolumn{4}{c|}{VBS}                                                                                            & \multicolumn{5}{c}{LSC}                                                                                             \\
            & 2021                        & 2022                       & 2023                       & 2024                        & 2020                        & 2021                        & 2022                       & 2023 & 2024                       \\ \hline
Teams       & 17                          & 16                         & 14                         & 32                          & 14                          & 17                          & 9                          & 14                     & 40    \\
Tasks       & 37                          & 34                         & 26                         & 34                          & 12                          & 23                          & 26                         & 30                  &   28     \\
Categories  & 3                           & 3                          & 4                          & 15                          & 1                           & 1                           & 3                          & 6                   & 6       \\
Submissions & 11811                       & 18124                      & 4452                       & 3246                        & 192                         & 399                         & 4373                       & 5055             &  7204        \\
Mode        & \multicolumn{1}{l}{virtual} & \multicolumn{1}{l}{hybrid} & \multicolumn{1}{l}{hybrid} & \multicolumn{1}{l|}{hybrid} & \multicolumn{1}{l}{virtual} & \multicolumn{1}{l}{virtual} & \multicolumn{1}{l}{hybrid} & \multicolumn{1}{l}{hybrid} & \multicolumn{1}{l}{physical}  \\
%Dataset        & \multicolumn{1}{l}{V3C1} & \multicolumn{1}{l}{V3C1 + V3C2} & \multicolumn{1}{l}{V3C1 + V3C2 + MVK} & \multicolumn{1}{l|}{V3C1 + V3C2 + MVK + VBSLHE	} & \multicolumn{1}{l}{LSC} & \multicolumn{1}{l}{LSC} & \multicolumn{1}{l}{LSC} & \multicolumn{1}{l}{LSC} \\
\hline
\end{tabular}
}

\end{table}

\pagebreak
\subsection{Localized Synchronous Evaluations}

Comparative experiments of retrieval systems can most easily be performed by having system instances next to each other and using them to solve the same tasks simultaneously in the same environment and under the same conditions.
Evaluating systems in such a setting eliminates many possible influences that would otherwise need to be considered.
This setting is intuitive and commonly used, from small-scale experiments to large-scale international evaluation campaigns, such as the VBS \cite{Lokovc2018:Influential,Schoeffmann2019:Video} or the LSC \cite{Gurrin2022:Introduction,Gurrin2023:Introduction,Gurrin2024:Introduction}. In 2023 and 2024, DRES served as a base infrastructure, powering both of these campaigns.
In each of these challenge instances, different evaluation systems, sometimes with multiple instances per system, were comparatively evaluated on a wide range of tasks, including known-item search tasks with different types of query descriptions, ad-hoc search tasks with live relevance assessment, and question answering tasks.

\subsection{Distributed Synchronous Evaluations}

While it can be desirable to have all systems participating in an evaluation in the same room to ensure a consistent environment for all of them, this is not always easy or even possible.
Especially from 2020 to 2022, various pandemic-related travel restrictions made conducting international evaluations at a common location impossible. During that time, the above-mentioned annual campaigns relied on DRES' support for distributed evaluations, such as VBS~\cite{DBLP:journals/ijmir/HellerGBG0LLMPR22,DBLP:journals/mms/LokocABDGMMNPRSSSKSVV23} and LSC~\cite{DBLP:journals/access/TranNDHSLPNKDJRYAATHSG23}, to have participants solve the same tasks simultaneously, but with each in a different location.
In order to do that, each participant had their own instance of the interface that would present the task information.
A simple synchronization mechanism ensures that all participants see the same information simultaneously (within a sufficiently small window of uncertainty), independently of network bandwidth and roundtrip time.
A common video call was used as a side-channel for exchanging non-technical information between all participants.

The possibility to run such evaluations in a distributed fashion not only serves as an emergency replacement for on-site experiments it also enables new kinds of evaluations that were not feasible previously.
Specifically, it lowers the cost and organizational overhead of scaling up comparative experiments to a larger number of participants.
Several such experiments, such as \cite{vitrivrVsSomhunter,schall2023interactive}, have been conducted so far, independently of the established campaigns.

\subsection{Distributed Asynchronous Evaluations}

When relaxing the locality requirements for comparative experiments, the next possible step is also to relax simultaneity.
This is possible due to DRES' support for asynchronous evaluations, where each participant solves tasks independently and not necessarily simultaneously as other participants.
This evaluation scheme enables to have a much larger number of participants, as there is no expensive synchronization overhead.
Such a setting has the additional benefit that participants can be recruited independently of each other, such as via a crowdsourcing platform.

We ran initial studies with over 200 participants recruited via prolific\footnote{\url{https://prolific.com}}, which is roughly an order of magnitude increase in participants when compared to the scenarios described above~\cite{willis2024task}.
Such settings, however, easily scale to a much larger number of participants.
This opens up possibilities for experiments that were previously not feasible.

\subsection{Distributed Non-interactive Evaluations}

While DRES' primary focus is on evaluation scenarios that require some interactivity, both the model described above and its implementation can also handle the more traditional non-interactive scenarios.
Since many existing campaigns rely on bespoke evaluation pipelines for every task, this has the potential of greatly simplifying existing processes and unifying procedures for evaluation participants.
Due to its modular architecture, DRES can rather straightforwardly be adapted to many tasks found in current campaigns such as TRECVid, CLEF, or MediaEval.

\section{Conclusion}
\label{sec:conclusion}
In this paper, we introduced a formal model for performance evaluation in multimedia retrieval and discussed its practical applicability.
We also presented DRES, an open-source evaluation infrastructure capable of supporting a broad range of retrieval evaluations.
Due to its flexible and modular architecture, DRES has already been used for a diverse range of retrieval experiments, including several international evaluation campaigns.
We also outlined further areas in which DRES might prove useful in the future.
With the introduction and the open-source release of DRES, we aim to support retrieval experiments and evaluations in related areas, reduce hurdles for large-scale experiments, and improve reproducibility.

\begin{acks}
    This work was partially supported by the Swiss National Science Foundation through project MediaGraph (contract no.\ 202125).
\end{acks}

\bibliographystyle{ACM-Reference-Format}
\bibliography{bibliography}

\end{document}

%% file: model-commands.tex
\newcommand{\powerseta}[1]{\ensuremath{\mathcal{P}(#1)}}
\newcommand{\paranthese}[1]{\ensuremath{\left(#1\right)}}
\newcommand{\tuple}[1]{\ensuremath{\left\langle#1\right\rangle}}
\newcommand{\set}[1]{\ensuremath{\left\lbrace#1\right\rbrace}}
\newcommand{\closedIntervala}[2]{\ensuremath{\left[#1,#2\right]}}

\def\realset{\ensuremath{\mathbb{R}}}
\def\realsetpos{\ensuremath{\mathbb{R}_{\geq 0}}}
\def\naturalset{\ensuremath{\mathbb{N}}}
\def\naturalsetz{\ensuremath{\mathbb{N}_0}}

\def\collection{\ensuremath{\Gamma}}
\def\mediaitem{\ensuremath{\gamma}}
\def\fragmentset{\ensuremath{\Omega}}
\def\fragment{\ensuremath{\omega}}
\def\desc{\ensuremath{\mathtt{desc}}}
\def\descset{\ensuremath{\mathtt{DESC}}}
\def\timeset{\ensuremath{T}}
\def\timep{\ensuremath{t}}
\def\reljudge{\ensuremath{\mathtt{rel}}}
\def\relundet{\ensuremath{\bullet}}
\def\answerscore{\ensuremath{r}}
\def\answer{\ensuremath{a}}
\def\answerset{\ensuremath{A}}
\def\agent{\ensuremath{p}}
\def\agentset{\ensuremath{P}}
\def\submissiontime{\ensuremath{t_s}}
\def\submission{\ensuremath{s}}
\def\submissionset{\ensuremath{S}}
\def\task{\ensuremath{z}}
\def\tasktemplate{\ensuremath{\hat{z}}}
\def\taskset{\ensuremath{Z}}
\def\tasktemplateset{\ensuremath{\hat{Z}}}
\def\timetaskstart{\ensuremath{t_{\task_s}}}
\def\timetaskend{\ensuremath{t_{\task_e}}}
\def\evalmode{\ensuremath{m}}
\def\evalmodeI{\ensuremath{\mathcal{I}}}
\def\evalmodeInt{\ensuremath{\mathcal{I}_i}}
\def\evalmodeNon{\ensuremath{\mathcal{I}_n}}
\def\evalmodeS{\ensuremath{\mathcal{S}}}
\def\evalmodeSyn{\ensuremath{\mathcal{S}_s}}
\def\evalmodeAsy{\ensuremath{\mathcal{S}_a}}
\def\eval{\ensuremath{e}}
\def\evaltemplate{\ensuremath{\hat{e}}}
\def\evalset{\ensuremath{E}}
\def\timeevalstart{\ensuremath{t_{\eval_s}}}
\def\timeevalend{\ensuremath{t_{\eval_e}}}
\def\activetask{\ensuremath{\mathtt{active}}}
\def\taskmetric{\ensuremath{f_{\mathit{task}}}}
\def\evalmetric{\ensuremath{f_{\mathit{evaluation}}}}
\def\auxparamset{\ensuremath{\Upsilon}}

%% file: figures/run-state-machine.tex
\begin{tikzpicture}[->,auto,node distance=3.5cm,initial text = ,]

    \node[initial,state,align=center]    (prepare_run)                          {Prepare\\Evaluation};
    \node[state,align=center]            (active_run) [right of=prepare_run]    {Evaluation\\Active};
    \node[state,accepting,align=center]  (ended_run)  [right of=active_run]     {Evaluation\\Ended};

    \node[state,align=center]    (init_task)    [below left=2cm and 1cm of prepare_run] {Create\\Task};
    \node[state,align=center]    (prepare_task) [right of=init_task]                    {Prepare\\Task};
    \node[state,align=center]    (active_task)  [right of=prepare_task]                 {Task\\Active};
    \node[state,accepting,align=center]    (ended_task)   [right of=active_task]                  {Task\\Ended};
    
    \path   (prepare_run)   edge[loop above,align=center]               node        {preparing\\content}   (prepare_run) 
            (prepare_run)   edge[align=center]                          node        {content\\ready}       (active_run)
            (active_run)    edge[loop above]                            node        {running}              (active_run)
            (active_run)    edge[align=center]                          node        {all tasks\\complete}  (ended_run)
          
            (init_task)     edge[align=center]                          node        {start task}            (prepare_task)
            (prepare_task)  edge[loop below,align=center]                            node        {waiting for\\participants}     (prepare_task)
            (prepare_task)  edge[align=center]                          node        {participants\\ready}          (active_task)
            (active_task)   edge[loop below,align=center]                            node        {accepting\\submissions} (active_task)
            (active_task)   edge[align=center]                             node        {task\\complete}             (ended_task)

            (active_run.200)    edge[dashed, bend right=15]                 node        {next task}             (init_task)
            (ended_task)    edge[dashed, bend right=15,align=center]    node        {task complete}        (active_run.340);

\end{tikzpicture}

%% file: figures/query-hints.tex
\begin{tikzpicture}[scale=.5,y=1cm,x=1cm,thick]
        % Timeline
        \draw[line width=1.2pt, ->, >=latex'] (-1,0) -- coordinate (x axis) (23,0) node[right] {$t$};
        % Vertical Axis
        \draw[line width=1.2pt] (0,-1) -- coordinate (y axis) (0, 13);
        
        % Channels
        \draw[line width=1.4pt] (-.2,2) -- (.2,2) node[left,align=right,xshift=-.5cm]{text};
        
        \draw[line width=1.4pt] (-.2,5) -- (.2,5) node[left,align=right,xshift=-.5cm]{image};
        
        \draw[line width=1.4pt] (-.2,8) -- (.2,8) node[left,align=right,xshift=-.5cm]{video};
        
        \draw[line width=1.4pt] (-.2,11) -- (.2,11) node[left,align=right,xshift=-.5cm]{audio};
        
        % Time Ticks
        \draw[line width=1.4pt] (1,-.2) -- (1,.2) node[below, align=center,yshift=-.5cm]{\small 00:00};
        \draw[line width=1.4pt] (3,-.2) -- (3,.2) node[below, align=center,yshift=-.5cm]{\small 00:30};
        \draw[line width=1.4pt] (5,-.2) -- (5,.2) node[below, align=center,yshift=-.5cm]{\small 01:00};
        \draw[line width=1.4pt] (7,-.2) -- (7,.2) node[below, align=center,yshift=-.5cm]{\small 01:30};
        \draw[line width=1.4pt] (9,-.2) -- (9,.2) node[below, align=center,yshift=-.5cm]{\small 02:00};
        \draw[line width=1.4pt] (11,-.2) -- (11,.2) node[below, align=center,yshift=-.5cm]{\small 02:30};
        \draw[line width=1.4pt] (13,-.2) -- (13,.2) node[below, align=center,yshift=-.5cm]{\small 03:00};
        \draw[line width=1.4pt] (15,-.2) -- (15,.2) node[below, align=center,yshift=-.5cm]{\small 03:30};
        \draw[line width=1.4pt] (17,-.2) -- (17,.2) node[below, align=center,yshift=-.5cm]{\small 04:00};
        \draw[line width=1.4pt] (19,-.2) -- (19,.2) node[below, align=center,yshift=-.5cm]{\small 04:30};
        \draw[line width=1.4pt] (21,-.2) -- (21,.2) node[below, align=center,yshift=-.5cm]{\small 05:00};
        
        % Query Hint 1 - Text 1
        %\node[rectangle,draw=black,fit={(1,1) (7,3)},inner sep=0pt, text centered] (t1) {\scriptsize \emph{``A wooden door being shut''}};
        \draw (1,1) rectangle (7,3);
        \node[inner sep=0pt,align=center,text width=2.5cm] at (4,2) {\scriptsize \emph{``A wooden door being shut''}};
        % Query Hint 3 - Text 2
        %\node[rectangle,draw=black,fit={(7,1) (13,3)},inner sep=0pt, text centered] (t2) {\scriptsize \emph{``A wooden door being shut by a person''}};
        \draw (7,1) rectangle (13,3);
        \node[inner sep=0pt,align=center,text width=4cm] at (10,2) {\scriptsize \emph{``A wooden door being shut by a person''}};
        
        % Query Hint 2 - Image 1
        %\node[rectangle,draw=black,fit={(3,4) (10,6)},inner sep=0pt, text centered,align=center] (i1) {Image of a door};
        \draw (3,4) rectangle (10,6);
        \node[inner sep=0pt,align=center,text width=4cm] at (6.5,5) {\small Image of a door};
        
        % Query Hint 4 - Audio 1
        %\node[rectangle,draw=black,fit={(7,10) (13,12)},inner sep=0pt,text centered,align=center] (a1) {\vspace{.2em}\small Bang of a door being shut};
        \draw (7,10) rectangle (13,12);
        \node[inner sep=0pt,align=center,text width=3.5cm] at (10,11) {\small Bang of a door being shut};
        
        % Query Hint 5 - Video 1
        %\node[rectangle,draw=black,fit={(13,7) (17,9)},inner sep=0pt,text centered,align=center] (v1) {Video segment};
        \draw (13,7) rectangle (21,9);
        \node[inner sep=0pt,align=center,text width=4cm] at (17,8) {\small Video segement};
\end{tikzpicture}

%% file: figures/dres-erd.tikz
\begin{tikzpicture}[
      font=\sffamily\scriptsize,
      on grid,
      node distance=5cm,
      %shorten >=1pt,
      auto,
      entity/.style={%
        rectangle,%
        inner sep=2mm,%
        outer sep=0pt,%
        draw=black,%
        text width=25mm,%
        align=center,%
        minimum height=12mm,%
        font=\sffamily\small\bfseries,%
      },%
      instanceof/.style={%
        -{Latex[open,length=4mm]},%
        very thick,%
      },%
      labelSB/.style={%
        sloped,%
        below,%
      },%
      labelHL/.style={%
        anchor=east,%
        text width=15mm,%
        align=center,%
      },%
      labelHB/.style={%
        anchor=north,%
        text width=15mm,%
        align=center,%
      },%
      cStart/.style={%
        very near start,swap,%
      },%
      cEnd/.style={%
        very near end,swap,%
      },%
      cStartT/.style={%
        very near start,%
      },%
      cEndT/.style={%
        very near end,%
      },%
      caStart/.style={%
        pos=.26,swap,%
      },%
      caEnd/.style={%
        pos=.76,swap,%
      },%
      templating/.style={%
        fill=green!20,%
      },%
      exec/.style={%
        fill=blue!20,%
      },%
      agents/.style={%
        fill=red!20,%
      },%
      testcol/.style={%
        fill=purple!20,%
      },%
      sub/.style={%
        fill=cyan!20,%
      },%
    ]

    \node[entity,agents] (user) {User};
    \node[entity,agents, below=of user] (team) {Team};

    \node[entity,templating, right=of user] (evaltemp) {Evaluation Template};
    \node[entity,templating, right=of evaltemp] (tasktemp) {Task Template};
    
    \node[entity,exec, right=of team] (eval) {Evaluation};
    \node[entity,exec, right=of eval] (task) {Task};

    \node[entity,sub, right=of task] (sub) {Submission};
    \node[entity,sub, above=2.5cm of sub] (answerset) {Answer Set};
    \node[entity,sub, above=of sub] (answer) {Answer};

    \node[entity,testcol, right=of answer] (mitem) {Media Item};
    \node[entity,testcol, below=of mitem] (mcol) {Media Collection};

    \draw[instanceof] (eval) -- node[labelHL,text width=10mm] {instance\\ of} (evaltemp);
    \draw[instanceof] (task) -- node[labelHL,text width=10mm] {instance\\ of} (tasktemp);

    \draw[] (team) -- node[labelHL,text width=10mm] {part of} (user)
        node[cStart] {1} node[cEnd] {N};

    \draw[] (team.north east) -- node[labelHL,text width=12mm] {participate\\ in} (evaltemp)
        node[cStart] {N} node[cEnd] {1};
    \draw[] (team) -- node[labelHB] {participate in} (eval)
        node[cStartT] {N}
        node[cEndT] {1};

    \draw[] (eval) -- node[labelHB] {comprises of} (task)
        node[cStartT] {1} node[cEndT] {N};
    \draw[] (evaltemp) -- node[labelHB] {comprises of} (tasktemp)
        node[cStartT] {1} node[cEndT] {N};

    \draw[] (task) -- node[labelHB] {has received} (sub)
        node[cStartT] {1} node[cEndT] {N};

    \draw[] (sub) -- node[labelHL,text width=12mm] {comprises\\ of} (answerset)
        node[caStart] {1} node[caEnd] {N};

    \draw[] (answerset) -- node[labelHL,text width=12mm] {comprises\\ of} (answer)
        node[caStart] {1} node[caEnd] {N};

    \draw[] (answer) -- node[labelHB] {references} (mitem)
        node[cStartT] {N} node[cEndT] {M};

    \draw[] (mitem) -- node[labelHL,text width=10mm] {part of} (mcol)
        node[cStart] {N} node[cEnd] {1};

    \coordinate[below=1cm of sub.south] (subsouth);
    \draw[] (team.south) -- node[caStart] {N} ++(0,-1) -- node[labelHB]{submitted by} (subsouth) -- (sub.south)
         node[caEnd] {1};

    % \draw[] ($1.2*(user.north west)$) rectangle ($1.2*(team.south east)$);

    \node[draw, dashed, fit=(user) (team)] (userbox) {};
    
    \node[draw, dotted, fit=(evaltemp) (tasktemp)] {};

    \node[draw, loosely dashed, fit=(eval) (task)]{};

    \node[draw, dash dot, fit=(sub) (answerset) (answer)]{};
    
    \node[draw, decorate,decoration=crosses, fit=(mitem) (mcol)] {};
\end{tikzpicture}